\documentclass[runningheads]{llncs}

\usepackage[british]{babel}
\usepackage[utf8]{inputenc}
\usepackage{subcaption}
\usepackage{booktabs}
\usepackage{xspace}
\usepackage{amsfonts}
\usepackage{mathtools}
\usepackage{multirow}
\usepackage{stmaryrd} %

\usepackage{microtype}
\usepackage{paralist}
\usepackage{csquotes}

\PassOptionsToPackage{hyphens}{url}
\usepackage[pdfborder={0 0 0}]{hyperref}
\makeatletter
\g@addto@macro{\UrlBreaks}{\UrlOrds}
\makeatother

\usepackage[style=lncs,maxbibnames=99]{biblatex}
\AtEveryBibitem{%
	\clearfield{isbn}
	\clearfield{doi}
	\clearfield{publisher}
	\clearfield{pages}
	\ifentrytype{book}{%
		\clearfield{url}%
	}{%
	}%
	\ifentrytype{article}{%
		\clearfield{url}%
	}{%
	}%
	\ifentrytype{inbook}{%
		\clearfield{url}%
	}{%
	}%
	\ifentrytype{inproceedings}{%
		\clearfield{url}%
	}{%
	}%
}
\DeclareDataInheritance{proceedings}{inproceedings}{
	\noinherit{editor}
	\noinherit{series}
	\noinherit{volume}
	\noinherit{publisher}
}
\DeclareDataInheritance{book}{inbook}{
	\noinherit{editor}
	\noinherit{series}
	\noinherit{volume}
}

\title{BOLD: A Benchmark for Linked Data User Agents and a Simulation Framework for Dynamic Linked Data Environments}

\author{%
	Tobias Käfer\inst{1}%
	\and%
	Victor Charpenay\inst{2}%
	\and%
	Andreas Harth\inst{3}%
}
\institute{%
	Karlsruhe Institute of Technology (KIT), Karlsruhe, Germany%
	\and%
	École des Mines Saint-Étienne and Univ.\ Clermont Auvergne, Saint-Étienne, France%
	\and%
	Friedrich-Alexander University (FAU), Nuremberg, Germany%
	\\\email{tobias.kaefer@kit.edu}, \email{victor.charpenay@emse.fr}, \email{andreas.harth@fau.de}}

\titlerunning{The BOLD Benchmark and Framework}

\begin{document}

\maketitle

\begin{abstract}
The paper presents the BOLD (Buildings on Linked Data) benchmark for Linked Data agents, next to the framework to simulate dynamic Linked Data environments, using which we built BOLD.
The BOLD benchmark instantiates the BOLD framework by providing a read-write Linked Data interface to a smart building with simulated time, occupancy movement and sensors and actuators around lighting.
On the Linked Data representation of this environment, agents carry out several specified tasks, such as controlling illumination.
The simulation environment provides means to check for the correct execution of the tasks and to measure the performance of agents.
We conduct measurements on Linked Data agents based on condition-action rules.
\end{abstract}

\begin{description}
\item[URL] \url{https://github.com/bold-benchmark/}
\item[License] GPLv3
\end{description}

\newcommand{\ie}{i.\,e.\@\xspace}
\newcommand{\eg}{e.\,g.\@\xspace}
\newcommand{\cf}{cf.\@\xspace}

\section{Introduction}
\label{sec:intro}

Technologies from the Semantic Web stack are nowadays the technologies of choice to provide interoperable interfaces to composite cyber-physical systems.
While controlling cyber-physical systems through such interfaces could benefit greatly from agent-oriented programming techniques, a Semantic Web environment has peculiar characteristics usually not found in agent environments.
Our goal is to contribute to the convergence of the research fields of agent-oriented programming and Semantic Web~\cite{DBLP:journals/dagstuhl-reports/BoissierCHR21,DBLP:journals/dagstuhl-reports/BoissierCHR23} by proposing a benchmark to evaluate agents that operate in Semantic Web environments.

Several initiatives illustrate the adoption of Semantic Web technologies for cyber-physical systems.
In manufacturing, the International Data Space uses Semantic Web technologies, with several use-cases around Industry 4.0~\cite{ids-ram}.
In building information management, several vocabularies have recently been introduced to represent data in buildings, such as Brick~\cite{DBLP:conf/sensys/BalajiBFGGHJKPA16} or the work of the Linked Building Data community group\footnote{\url{https://www.w3.org/community/lbd/}} at the World Wide Web Consortium (W3C).
Works orthogonal to sub-fields of cyber-physical systems include W3C Recommendation such as the Semantic Sensor Network ontology (SSN)~\cite{vocab-ssn2017}, a widely adopted vocabulary to describe sensor and actuator data, and
the Web of Things (WoT) effort's WoT Architecture~\cite{wot-arch} and Thing Description~\cite{wot-td}, which allow to access sensor and actuator data from the devices.

The Linked Data~\cite{rwld} environment underlying those efforts is characterized by:
\begin{compactenum}
  \item Hypermedia, i.\,e.\@ the possibility of following links, allowing agents to discover new system components and possible actions to perform on them,
  \item Semantic alignments, allowing agents to bridge between terminologies of different system components, and
  \item RESTfulness, thus agents can read/write information from/to various systems
\end{compactenum}

Most attention in Linked Data research has been given to perfecting the environment, that is, storing and serving data at scale by servers \cite{DBLP:conf/icde/SchmidtHLP09,DBLP:conf/semweb/HartigBF09,DBLP:conf/otm/KeppmannMH17}.
Comparatively little work has been done on Linked Data agents, the flip side of the same coin.
This gap~\cite{DBLP:conf/atal/CiorteaMGBRZ19} could be filled in collaboration with the the agent research community, which has been developing Multi-Agent System (MAS) architectures for manufacturing and building automation for decades~\cite{van_brussel_reference_1998,DBLP:conf/icmas/HubermanC95,hagras2004creating}.

However, these architectures feature agents that are physically situated in their environment.
That is, agents are physically constrained in perceiving and acting~\cite{ferber1996influences}. %
In contrast, %
Linked Data agents are never strictly situated:
\begin{compactenum}
  \item Linked Data servers make sensor data visible to all agents without distinction. Linked Data agents must restrict the scope of their perception to data relevant to their individual goal, e.g.\ by following certain links only.
  \item Servers expose and consume symbolic representations of the physical environment. %
  Different servers use different vocabularies that can be semantically aligned using mappings. Linked Data agents must interpret such alignments. %
  \item All potential actions may be exposed to Linked Data agents simultaneously. To take full advantage of parallel actions and avoid conflicts, Linked Data agents must restrict the scope of their actions.
\end{compactenum}
Those points influence all three high-level phases of an agent's cognitive loop: sensing, reasoning, and action.
As a result, the existing benchmarks that already exist for MAS and that assume situated agents (e.g. \cite{wilensky_introduction_2015}) are not applicable. %

In this paper, we introduce a benchmark to specifically evaluate and compare Linked Data agents, their architectures and runtimes.
Our benchmark, the Building on Linked Data (BOLD) benchmark, is based on a real-world system configuration in the building automation domain.
The key contributions are: %
\begin{compactitem}
  \item A formal method for modelling dynamic Linked Data environments, and evaluating agent performances over such environments;
  \item A corresponding execution framework;
  \item An instantiation of our method and framework:
\begin{compactitem}
\item A dynamic Linked Data model of a building (building 3 of IBM Research Dublin) with simulated occupancy and lighting; and
  \item A set of corresponding agent evaluation tasks with increasing complexity %
\end{compactitem}
\end{compactitem}

This paper is structured as follows:
in Sec.~\ref{sec:example}, we provide a small example to illustrate the overall system; in Sec.~\ref{sec:preliminaries}, we provide preliminaries around the benchmark environment; subsequently, in Sec.~\ref{sec:dataset} and \ref{sec:simulation-environment}, we derive the design of the BOLD environment; then, in Sec.~\ref{sec:tasks}, we present tasks for user agents to carry out; next, in Sec.~\ref{sec:evaluation}, we report on applications of BOLD; in Sec.~\ref{sec:related-work}, we survey related work; and last, in Sec.~\ref{sec:conclusion}, we summarise our work.

\section{Example}\label{sec:example}

We now use an example to illustrate the main aspects of the BOLD benchmarking system%
\footnote{We assume the reader has a basic understanding of HTTP, RDF and SPARQL. The interested reader can find more details on these technologies \eg in \cite{DBLP:books/crc/linked2014}. We follow the prefix practices recorded in \url{http://prefix.cc/}, except \nolinkurl{:} as short for \nolinkurl{http://localhost:8080/}, \nolinkurl{brick:} as short for \url{http://buildsys.org/ontologies/Brick\#}, and \nolinkurl{bf:} as short for \url{http://buildsys.org/ontologies/BrickFrame\#}.}.
We first characterize the resources the server makes available as part of the environment, in the example the Coffee Dock room of IBM Research Dublin's building 3.
Next, we give an overview of how an agent carries out a simple task, in the example switching on the light in the Coffee Dock room.
Finally, we illustrate how the server tracks the agent's progress over time.

The resources relevant for the example are: The Coffee Dock Room, \nolinkurl{:Room_CoffeeDesk}, the room's lighting system, \texttt{:Lighting\_System\_42GFL\-Coffee\-Dock}, and the lighting system's on/off property, \texttt{:property-Lighting\_System\_42GFL\-Coffee\-Dock\#it}.
The description of the building uses the Brick ontology~\cite{DBLP:conf/sensys/BalajiBFGGHJKPA16}, plus terms from SOSA/SSN~\cite{vocab-ssn2017} to describe sensors and actuators, in particular the properties of sensing and actuation systems that agents can read and write.

For an agent to switch on the light in the Coffee Dock Room, the agent has to first discover the URI of the light.
Thus, the agent performs a HTTP \texttt{GET} request on \nolinkurl{:Room_CoffeeDesk}, which leads to a response in RDF that contains a link to the lighting system.
The agent next performs a traversal of this link and dereferences the IRI of the lighting system.
The thus obtained graph contains a link to the on/off property of the lighting system and traverses.
Finally, to achieve the task, the agent must write the value \texttt{on} to the lighting system property, which is done by an HTTP \texttt{PUT} request with a suitable RDF payload.

To test whether an agent succeeds in a task, each task is defined by `faults' in the environment that the agent must fix, in MAS terms, a norm violation.
In the simple task of turning on the light, there is one fault to fix, defined as the Coffee Dock lighting system having the property value \texttt{off}, expressed as a SPARQL \texttt{ASK} query.
An agent has succeeded in the task if the query evaluates to false.

The simple task of turning on the light illustrates the basic case which just needs a single agent loop and for which a simple success metric suffices.
The benchmark also includes tasks in which the agent has to loop and act continuously, which requires a more elaborate server setup (eg.\ to provide dynamic data), together with a fault rate metric to capture the success of agents appropriately.

\newcommand{\nop}[1]{}

\nop{
The fault for this simple task translates into a SPARQL query as follows:

\begin{footnotesize}
\begin{verbatim}
BASE <http://localhost:8080/>
ASK
FROM <property-Lighting_System_42GFLCoffeeDock>
WHERE {
   <property-Lighting_System_42GFLCoffeeDock#it> rdf:value "on" .
}
\end{verbatim}
\end{footnotesize}

The BOLD server discretizes time into timeslots of equal length and runs this query at the end of every timeslot.
The server then records the performance of the agent as the {time of first success}, that is, the lowest timeslot
index at which the query evaluates to false. The time of first success shown on Fig.~\ref{fig:coffee-dock-seq} is
2.

Next, consider the continuous task of switching lights off when no occupant is detected in the Coffee Dock room.
The task requires an occupancy sensor. From the earlier retrieval of the room's description, the agent now knows
that a description of the room's occupancy sensor is available under the following URI:

\begin{footnotesize}
\begin{verbatim}
http://localhost:8080/Occupancy_Sensor_42GFLCoffeeDock
\end{verbatim}
\end{footnotesize}

Performing link traversal again, the agent then discovers a new property for occupancy, whose value it can retrieve.
When considering occupancy, the task requires the server to simulate changes in the environment, such that the occupancy value of the Coffee Dock room eventually turns false.
In BOLD, we specify changes via SPARQL updates. %
For example, consider a SPARQL update that simulates the fact that at 8am on a particular day, someone entered the Coffee Dock room:

\begin{footnotesize}
\begin{verbatim}
BASE <http://localhost:8080/>
DELETE { 
   GRAPH <property-Occupancy_Sensor_42GFLCoffeeDock> {
      <property-Occupancy_Sensor_42GFLCoffeeDock#it> rdf:value "off"
   }
} INSERT {
   GRAPH <property-Occupancy_Sensor_42GFLCoffeeDock> {
      <property-Occupancy_Sensor_42GFLCoffeeDock#it> rdf:value "on"
   }
} WHERE {
   <sim> sim:currentTime ?time .
   ?time time:inXSDDateTimeStamp "2020-05-22T08:00:00-12:00"
}
\end{verbatim}
\end{footnotesize}

The BOLD environment server would internally execute the update and expose the result to the agent, so that the agent can act accordingly.
A realistic simulation would generate several state changes over a single day, \eg for lunch breaks. A task that includes occupancy thus requires more than a single update in the environment.
The agent must continuously monitor the environment for changes.

To verify how well an agent has carried out the continuous task we introduce here, we can use the notion of faults as defined earlier but restricted to cases where a building occupant is detected in the room.
Such a restriction requires that an agent both monitors the room's luminaire and its occupancy sensor.
The restricted fault can be specified by the following query:

\begin{footnotesize}
\begin{verbatim}
BASE <http://localhost:8080/>
ASK
FROM <property-Lighting_System_42GFLCoffeeDock>
FROM <property-Occupancy_Sensor_42GFLCoffeeDock>
WHERE {
   <property-Occupancy_Sensor_42GFLCoffeeDock#it> rdf:value "off" .
   <property-Lighting_System_42GFLCoffeeDock#it> rdf:value "on" .
}
\end{verbatim}
\end{footnotesize}

For what we call `continuous-loop' tasks, which require continuous monitoring by the agent, the time of first success is not a satisfactory metric anymore.
Instead, a \emph{fault rate} metric is considered: if the environment server evaluates the above query at fixed intervals, fault rate is the ratio of times such that the query evaluates to true over the duration of the total duration of a simulation run.
On Fig.~\ref{fig:coffee-dock-seq}, if we assume the simulation stops at index 5 and if we further assume that an occupant entered and left the Coffee Dock room at timeslot 3, the success rate for our agent equals (6-2)/6, 75\%.

}

\section{Formal Definitions and Execution Framework}
\label{sec:preliminaries}

We now introduce a series of concepts to formally define the tasks and metrics used in BOLD.
We build on~\cite{DBLP:conf/emas/CharpenayKH21} for the definitions.
A task is defined by a set of faults that agents have to fix.
Such tasks correspond to the MAS concept of norms, i.\,e.\ constraints to be satisfied.
The evaluation of a task depends on a simulation environment.
We define the simulation environment and metrics for faults and performance.
Last, we introduce a corresponding execution framework.

\subsection{Dataset and Simulation Run}
\label{sec:sim-run-preliminaires}

We use the usual abstract notation for RDF. $I$, $B$, $L$ are disjoint sets of IRIs, blank nodes and literals.
An RDF quad is an instance of $I \cup B \times I \times I \cup B \cup L \times I$ and an RDF dataset is a set of RDF quads. $D$ denotes the set of RDF datasets.

\begin{definition}[simulation run, simulation environment]
A simulation run is a finite sequence of datasets $\langle d_{0}, d_{1}, \ldots, d_{k} \rangle$, $k \in \mathbb{N}$.
A simulation environment can be defined as the pair $\langle d, u \rangle$,
where $d \in D$ and $u$ is some update function $u : D \mapsto D^n$, such that for all possible simulation run $\langle d_{0}, d_{1}, \ldots, d_{k} \rangle$ of the environment the following holds: $d_0 = d$ and $ \forall t < k: d_{t+1} \in u(d_{t})$
\end{definition}

The update function $u$ corresponds to a (set of) SPARQL \texttt{UPDATE}(s). %
A simulation environment can thus be seen as a transition system that generates a set of simulation runs (of arbitrary size).
SPARQL updates may include non-deterministic function calls. The output of $u$ is therefore a set of RDF datasets.%
We assume $n$ to be finite, so that SPARQL updates are only allowed to include a finite set of binary decisions of the form \texttt{rand() < ?threshold}.

The above definition only defines `dry runs', during which the environment is updated automatically without agent intervention.
Between each environmental update, though, agents can act on the environment by updating a dataset $d_i$. Linked Data interfaces only allow a limited set of operations
on Web resources (\texttt{PUT}, \texttt{POST}, \texttt{DELETE}), as formalized below (\texttt{GET} corresponds to $id$).

\begin{definition}[agent operation]
Let $\delta : D \times D \mapsto D$ be the function calculating the symmetric difference between two datasets: $\delta(d, d') = (d \cup d') \setminus (d \cap d')$.
Moreover, let $\pi$ be a projection function such that $\pi(d)$ is the set of resource IRIs (graph names) appearing in $d$.
A function $op : D \mapsto D$ is called an agent operation if $\pi(\delta(d, op(d)))$ is a singleton. The identity function $id(d) = d$ is an agent operation. On top, we use the following terminology:
\begin{compactitem}
    \item if $\pi(d) \subset \pi(op(d))$, $op$ is a 'create' operation
    \item if $\pi(op(d)) \subset \pi(d)$, $op$ is a 'delete' operation
    \item otherwise, $\pi(d) = \pi(op(d))$ and $op$ is a 'replace' operation
\end{compactitem}
\end{definition}

A full benchmark run is thus the interleaving of agent operations and environmental updates, obtained through composition, see $\circ$ in the following definition.

\begin{definition}[benchmark run]
Let $\langle d, u \rangle$ be a simulation environment.
A benchmark run $\rho$ is a finite sequence $\langle d_{0}, d_{1}, \ldots, d_{k} \rangle$, such that:
(1) $d_0 = d$, and (2) $\forall t < k$, there is a finite sequence of agent operations $\langle op_1, op_2, \ldots op_l \rangle$, such that $d_{t+1} \in op_1 \circ op_2 \circ \ldots op_l \circ u(d_t)$
\end{definition}

\subsection{Faults and Performance Metrics}
\label{sec:tasks-preliminaires}

\begin{definition}[fault sequence]
A fault sequence $\gamma$ is a finite sequence of datasets $\langle d_{0}, d_{1}, \ldots, d_{l} \rangle$, $l \in \mathbb{N}$.
A run $\rho = \langle d_{0}, d_{1}, \ldots, d_{k} \rangle$ matches a fault sequence $\gamma = \langle d'_{0}, d'_{1}, \ldots, d'_{l} \rangle$ at time $t$ if a) $l \leq t \leq k$ and b) $\forall i | t - l \leq i \leq t: d_i \cap d'_i = d'_i$
\end{definition}

Fault sequences can be defined in terms of sequences of SPARQL queries.
Several fault sequences may occur in the environment at a given time. In the following, we denote $\Gamma$ the (possibly infinite) set of fault sequences defined for a task.
We also denote $\Gamma_{\rho,t}$ the set of all $\gamma \in \Gamma$ such that $\rho$ matches $\gamma$ at time $t$.

Every task is associated with potential fault sequences in the environment that agents must fix as quickly as possible.
In the Coffee Dock room example, the SPARQL \texttt{ASK} query would be associated to one task (as a fault sequence of size one).
Given a run $\rho$ and a a set of fault sequences $\Gamma$ (defining a task), one can specify performance metrics to evaluate $\rho$ against $\Gamma$.

\begin{definition}[success rate]
\label{def:success-rate}
Let $\rho$ be a benchmark run of size $k$ and $\Gamma$ a set of fault sequences of equal size $l$.
The fault rate is the number of times at which $\rho$ matches some $\gamma \in \Gamma$ divided by the length of $\rho$:

\begin{equation}
    \frac{| \{ t \; | \; \Gamma_{\rho,t} \neq \emptyset \} |}{k - l}
\end{equation}
\end{definition}

When the task requires a single action loop (see Sec.~\ref{sec:tasks}), the fault rate gives an indication of `how fast' an agent is at solving the task.
When the task requires continuous action, the metric gives an indication of `how efficient' an agent is over the whole duration of the task.
Counting faults gives an indication of `how close' an agent has been to solving the task on average.
\begin{definition}[average fault count]
\label{def:avg-closeness}
Let $\rho$ be a benchmark run of size $k$ and $\Gamma$ be a set of fault sequences of equal size $l$. The average fault count is the sum of $\Gamma_{\rho,t}$ over $\rho$:
\begin{equation}
    \frac{\sum_{l \leq t \leq k} |\Gamma_{\rho,t}|}{| \{ t \; | \; \Gamma_{\rho,t} \neq \emptyset \} |}
\end{equation}
\end{definition}

We do not only want to compare different agent architectures, but also the same architecture \textit{across} tasks.
To this end, we further define a task-independent fault count for a run that is normalized against a certain dry run.

\begin{definition}[normalized fault count]
\label{def:nfcount}
Let $\langle u, d \rangle$ be a simulation environment.
Let $\rho$ and $\overline{\rho}$, both of size $k$ and both starting with $d$, be respectively a benchmark run and a simulation run in the environment.
For all $\overline{d_t}$ of $\overline{\rho}$ ($t < k$), it holds that $\overline{d_{t+1}} \in u(\overline{d_t})$.
The ratio

\begin{equation}
    \frac{\sum_{l \leq t \leq k} |\Gamma_{\rho,t}|}{\sum_{l \leq t \leq k} |\Gamma_{\overline{\rho},t}|}
\end{equation}
is a normalized fault count if for all $d_t$ of $\rho$ ($t < k$): $d_{t+1} \cap \overline{d_{t+1}} = u(d_t) \cap u(\overline{d_t})$.
\end{definition}

The above definition ensures that $\rho$ and $\overline{\rho}$ are comparable in how non-deterministic functions behave during their execution, regardless of agent operations.
In practice, it is enough to use a pseudo-random number generator with the same seed for all runs that must be compared. The resulting normalized fault count then measures
closeness to success for a task regardless of how many potential faults the task defines.

While the metrics defined so far relate to minimizing faults, they do not penalize agents that would perform unnecessary operations to reach their goal. To take this aspect into account, we finally introduce read/write ratio as a metric.

\begin{definition}[read/write ratio]
Let $\rho$ be a benchmark run of size $k$.
Let, $\forall t < k$, be a sequence $\langle op_{t,1}, op_{t,2}, \ldots op_{t,l_t} \rangle$, such that $d_{t+1} \in op_{t,1} \circ op_{t,2} \circ \ldots op_{t,l_t} \circ u(d_t)$.
Each operation sequence at time $t$ may include multiple occurrences of $id$, capturing read operations performed by agents. We denote $i_t$ the number of occurences of $id$ at time $t$.
We thus define the read/write ratio for $\rho$ as follows:

\begin{equation}
    \frac{\sum_{t < k} i_t}{\sum_{t < k} l_t - i_t}
\end{equation}
\end{definition}

\subsection{Execution Framework for Dynamic Linked Data Simulations}
Our execution framework is based on Eclipse RDF4J (formerly: Sesame~\cite{DBLP:conf/semweb/BroekstraKH02}) and controls the simulation, maintains the dataset, and executes SPARQL \texttt{UPDATE} / \texttt{ASK} queries, using which we implemented the simulation's dynamics / fault checks.
We wrote SPARQL user-defined functions to facilitate this simulation.
The REST interface is built on Java Servlets and the SPARQL Graph Store Protocol, and runs on Eclipse Jetty.
Using \texttt{.properties} files, people can register datasets, initialisation queries, simulation updates, and/or fault checks.
This way, people can build a repeatable standard-based dynamic Linked Data environment.
\section{Dataset}
\label{sec:dataset}

Our benchmark builds on Balaji et al.'s datasets that describe real buildings, which they published using their Brick ontology~\cite{DBLP:conf/sensys/BalajiBFGGHJKPA16}.
Of the datasets in~\cite{DBLP:conf/sensys/BalajiBFGGHJKPA16}, we chose building 3 of IBM Research Dublin, as it has diverse building systems with discrete and continuous state, including lighting, and built a Linked Data version.

\subsection{The Brick Ontology}

The Brick ontology is a simple model to represent buildings and their associated automation systems~\cite{DBLP:conf/sensys/BalajiBFGGHJKPA16}.
We base our considerations on version 1.0.0 of the Brick ontology, as this is the last version under which building 3 has been published.

\begin{figure}[t]
\centering
\includegraphics[scale=0.3]{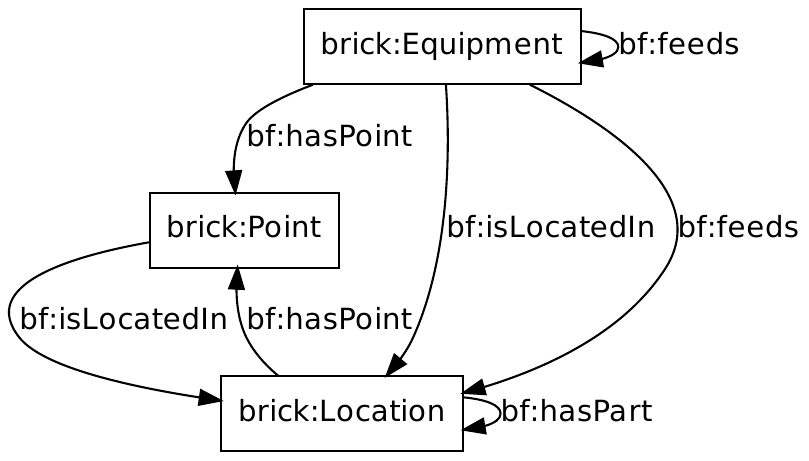}
\caption{\label{fig:brick-prop}The Brick Ontology's RDF Properties as UML class diagram. UML class associations between RDFS classes depict \href{http://www.w3.org/2000/01/rdf-schema\#domain}{domains} and \href{http://www.w3.org/2000/01/rdf-schema\#range}{ranges} of the \href{http://www.w3.org/1999/02/22-rdf-syntax-ns\#Property}{properties}.}
\end{figure}

The main classes we use from the Brick ontology are~\cite{brick-leaflet}, see Figure~\ref{fig:brick-prop}:
\begin{compactitem}
    \item \nolinkurl{brick:Equipment} refers to any technical equipment as part of some building automation system, including fire safety systems, HVAC (heating, ventilation and air conditioning), and lighting systems
    \item \nolinkurl{brick:Point} refers to the data points (sensors, commands, set points) to monitor and control building automation systems
    \item \nolinkurl{brick:Location} refers to any location inside a building (such as floors and rooms) that is of relevance for automation
\end{compactitem}
The Brick ontology also defines properties:
\begin{compactitem}
\item \nolinkurl{bf:hasPoint} can relate locations and equipment to a \nolinkurl{brick:Point}.
\item \nolinkurl{bf:hasPart} and its inverse \nolinkurl{bf:isPartOf}, are transitive properties to define equipment and locations in a hierarchical fashion
\item \nolinkurl{bf:isLocatedIn}, a transitive property to specify the location of some equipment or a data point
\item \nolinkurl{bf:feeds}, a property to specify what locations are covered (fed) by some automation system, either directly or indirectly via other systems.
\end{compactitem}

\subsection{Description of Building 3 in Brick}
\label{sec:ibm_b3}

Building 3 of IBM Research Dublin is a two-storey office building.
The building has a rectangular ground plot with the long side oriented from north to south, \ie most rooms are oriented eastwards or westwards.
In the building description\footnote{\url{https://github.com/BuildSysUniformMetadata/GroundTruth/blob/2e48662/building_instances/IBM_B3.ttl}},
we see a subdivision into floors and wings, to which the rooms are assigned using \nolinkurl{bf:hasPart} relationships.
The rooms and wings have \nolinkurl{brick:Lighting_Systems} assigned.
Such a light system comes in different forms: Some come with a \nolinkurl{brick:Occupancy_Sensor} that determines whether there are people in its vicinity, others with a \nolinkurl{brick:Luminance_Command}, in other words, a switch to control the lights, and others with a \nolinkurl{brick:Luminance_Sensor} that we consider to be triggered by daylight.
We provide basic statistics in Table~\ref{tab:b3-basics}.

\subsection{Building 3 in Read-Write Linked Data}
\label{sec:ibm_b3-ext}

The description of building 3 as found online is one single static RDF graph with no dereferenceable URIs and no entity with a temporal extent (such as sensor measurements).
To make it suitable for a Read-Write Linked Data benchmark, we extended this monolothic RDF graph in two ways.
First, we scoped every triple in the graph with a dereferenceable resource IRI, such that the result is a proper RDF dataset, as defined in Sec.~\ref{sec:preliminaries}.
Second, we extended the definition of the building's data points, in order to provide resources that change over time.

\subsubsection{Resource Partitioning}

As our aim is to provide a benchmark for Web agents, resource IRIs must be dereferenceable.
Morever, we assume that cyber-physical systems with read-write Linked Data interfaces will be composed of large amounts of resources of small size (individually exposed by connected devices).
To match this assumption, we partitioned the original RDF graph into smaller RDF graphs, each providing information about one room, one floor, one data point, etc\@.
Specifically, we created an RDF dataset such that from every triple $\langle s, p, o \rangle$ from the graph, we derived the quads $\langle s, p, o, s \rangle$ and $\langle s, p, o, o \rangle$.
We thus derived about 3k dereferenceable IRIs from 25k triples, with graphs of 5-50 triples (see Table~\ref{tab:b3-basics}).

\subsubsection{Time-varying Resource Definition}

\begin{figure}[t]
\centering
\includegraphics[scale=0.3]{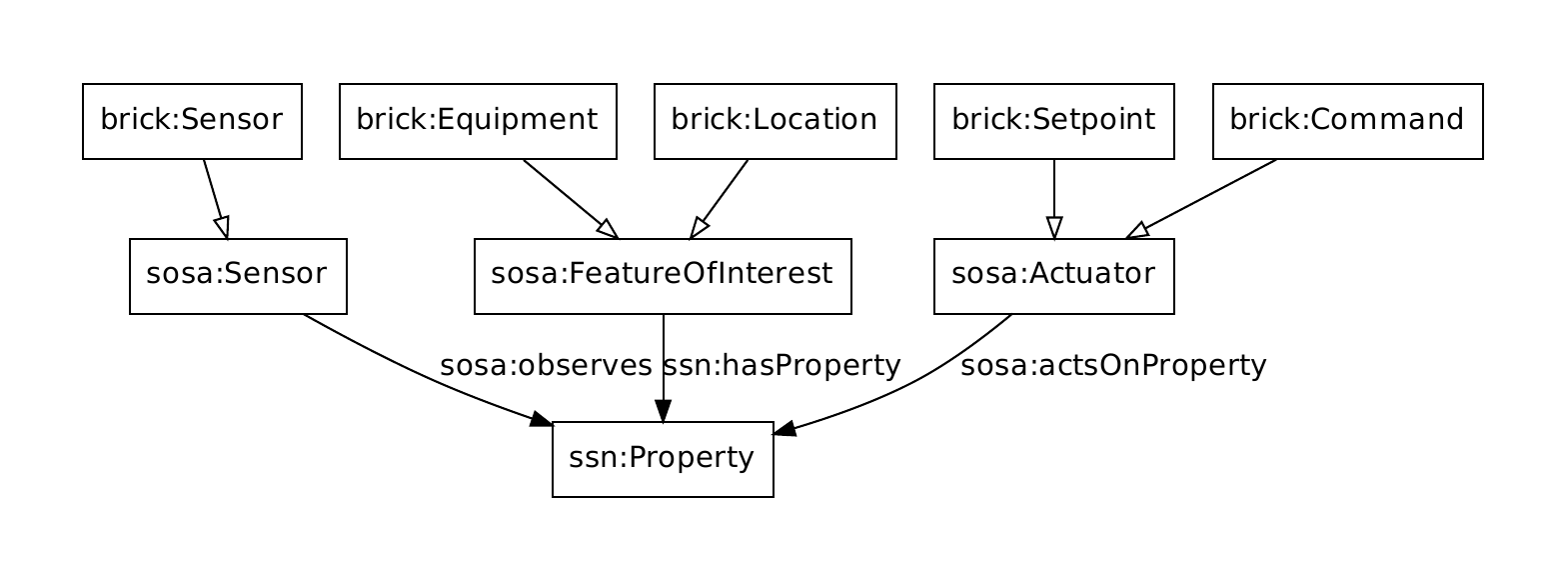}
\caption{\label{fig:brick-ssn-mapping}Mapping from the Brick ontology to the SSN ontology in the form of an UML Class Diagram. See Figure~\ref{fig:brick-prop} for an explanation of our notation.}
\end{figure}

Although the original RDF graph includes resources like sensors and actuators, it does not provide means to expose actual sensor measurements (such as occupancy or an illuminance value) nor does it provide means to expose the state of actuators (such as the status of a light switch).
To complete the dataset on which BOLD is defined, we included time-varying and writable resources for such sensor and actuator data.
To that end, we aligned the Brick ontology with SOSA/SSN as illustrated on Fig.~\ref{fig:brick-ssn-mapping}. We then extended our dataset with resources defined as instances of ssn:Property.
For all data points of the building 3 lighting systems, we created a property resource similar to the example payload of Sec.~\ref{sec:example}.
We only consider lighting systems since all tasks of BOLD are defined according to a lighting scenario (see Sec.~\ref{sec:tasks}).
As a result, we added $>500$ dereferenceable IRIs to our Linked Data benchmark, see Table~\ref{tab:b3-basics}.
Section~\ref{sec:simulation-environment} describes how these resources vary over time.

\begin{table}[tb]
\caption{\label{tab:b3-basics}Basic counts for Building~3 and the benchmark.}
\centering
    \begin{tabular}{lr}
\toprule
Rooms                        & 281 \\
\quad with occupancy sensors   & 66 \\
\quad with luminance commands  & 38 \\
\quad with luminance sensors   &  20 \\
Floors                       &   2 \\
Wings                        &   3 \\
&\\
\bottomrule
\end{tabular}
    \begin{tabular}{lr}
	\toprule
Lighting systems             & 278 \\
\quad with occupancy sensors   & 156 \\
\quad with luminance commands  & 105 \\
\quad with luminance sensors   &  48 \\
Triples in IBM\_B3.ttl       & 24\,947 \\
Resources IRIs               &  3\,281 \\
Dynamic resources            &     551 \\
\bottomrule
    \end{tabular}
\end{table}

\section{Simulation Environment}
\label{sec:simulation-environment}

The simulation environment is defined by an initial dataset and an update function that non-deterministically updates the dataset exposed to agents.

The dataset presented in Sec. \ref{sec:ibm_b3} serves as initial dataset in the simulation. We now present how datasets are updated by the simulation environment and how a dataset is exposed to agents.

\subsection{Time}

The primary role of the simulation environment is to increment time.
To this end, a timer is triggered at fixed intervals.
The timer increments a number and progresses an RDF description of the simulated time (in OWL Time~\cite{owl-time}).
Time is made available to agents under a \url{:sim} resource, to which agents can also send a HTTP \texttt{PUT} request to start a simulation run.

\nop{
Time is made available under a \texttt{<sim>} resource, as in the following example\footnote{The \texttt{sim:} prefix denotes a namespace that is exposed locally to agents.}:

\begin{small}
\begin{verbatim}
<sim> sim:currentIteration 12 ;
      :currentTime [
          time:inXSDDateTimeStamp
            "2020-05-23T00:12:00-12:00"^^xsd:dateTime ;
          time:inDateTime [ time:year 2020 ; time:month 05 ] # ...
      ] .
\end{verbatim}
\end{small}

This resource can also be updated by agents to start a simulation run. A \texttt{PUT} request to \texttt{<sim>} with the following payload will start a simulation run:

\begin{small}
\begin{verbatim}
<sim> sim:initialTime
        "2020-05-22T23:59:00-12:00"^^xsd:dateTime ;
      sim:timeslotDuration 60000 ; # in milliseconds
      sim:iterations 720 . # 12 hours
\end{verbatim}
\end{small}
}

Time, as exposed to agents, corresponds to simulated time. Real execution time can be parameterized on the simulation server as the period of a timeslot between two successive environmental updates.
However, care must be taken that all updates can be computed in less than the duration of a timeslot, to prevent time drift.
Our implementation of the BOLD simulation server can compute all environmental updates in $<100$\,ms (avg.\ 67$\pm$28\,ms for tasks w/sunlight and occupancy, see Sec.~\ref{sec:tasks}).
We obtained all results (Sec.~\ref{sec:evaluation}) with this configuration.

\subsection{Sunlight and Occupancy}

Some BOLD tasks rely on sensor measurements that result from two physical processes to simulate:
the illumination of the building by the sun and the movements of occupants in the building.
Although they take different parameters into account, the simulation of both processes have the following characteristics:
\begin{compactenum}
    \item non-determinism of transitions so that agents may not optimize using out-of-band information
    \item predictable evolution to allow for model-based agent optimization
\end{compactenum}

\begin{figure}[bt]
    \centering
    \begin{subfigure}[b]{0.45\textwidth}
        \includegraphics[width=.9\textwidth]{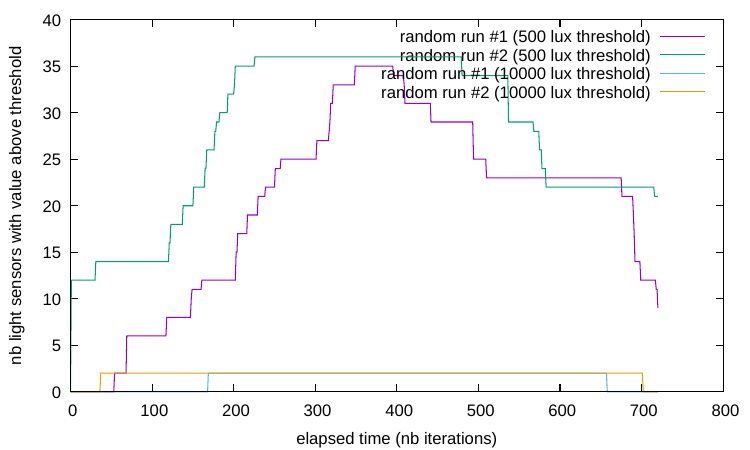}
        \caption{Sunlight}
        \label{fig:sim-luminance}
    \end{subfigure}
    \begin{subfigure}[b]{0.45\textwidth}
        \includegraphics[width=.9\textwidth]{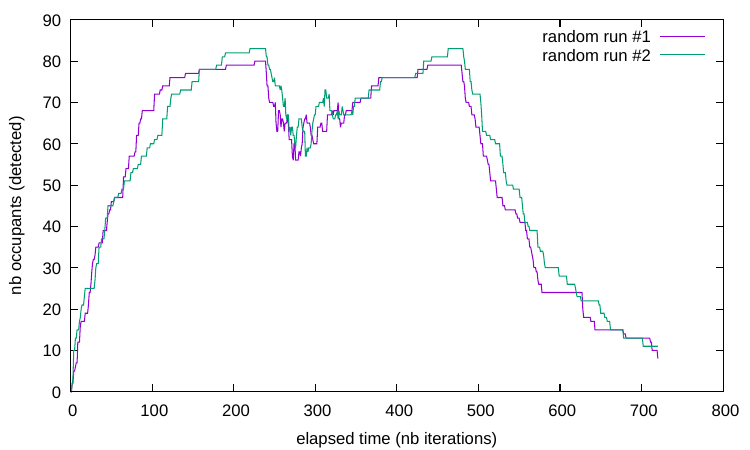}
        \caption{Occupancy}
        \label{fig:sim-occupants}
    \end{subfigure}
    \caption{Example of (plain) random runs for sunlight and occupancy}
    \label{fig:sim-examples}
\end{figure}

An example is provided on Fig.~\ref{fig:sim-examples}. In the case of sunlight, outside illuminance (measured in lux) evolves quadratically over time, as the effect of the sun rising (6am), reaching a zenith and then setting (9pm).
Illuminance at zenith (\ie the maximum value) equals 40k lux. 
A cloud coverage factor is applied to that baseline, as follows: a random coverage in $[0,1]$ is generated at sunrise and at sunset, then coverage evolves linearly between these two values.
Coverage either increases or decreases during the day (illustrated by run 1 and 2 resp.\ in Fig.~\ref{fig:sim-luminance}).

Inside a building, the intensity of light that is incident on walls and floors is significantly lower than sunlight as measured outside.
In BOLD, we randomly assign to each room an `occlusion factor' in $[0.5,1]/10$ to account for this phenomenon. As a result, the maximum illuminance inside the building is 4,000\,lux.

To simulate the movements of occupants, time is divided in 4 periods, see Fig.~\ref{fig:sim-occupants}:
At night, the building has no occupant. From 8am, occupants start arriving at random times and occupy one room each.
After 12pm, occupants leave the building for lunch, about 1\,h, after which they may come back for another 4\,h of work. Starting from 4pm, occupants start leaving the building.

\subsection{Linked Data Interface}

Agents can access and control the latest state of the simulation via a Linked Data interface that complies to the SPARQL Graph Store Protocol (with direct addressing) \cite{gsp2013}.
That is, for a dataset $d_i$, agents can send HTTP requests to any resource IRI $r \in \pi(d_i)$. All resources in the dataset are readable except the default graph, which holds triples that are used to compute the simulation's update function (e.g. triples stating the existence of occupants and what room is their workplace).
On top, instances of \texttt{so\-sa:Ac\-tua\-table\-Pro\-per\-ty} are writable.

Our implementation is multi-threaded: one thread triggers environmental updates at the end of each timeslot while a thread pool concurrently accesses the dataset to carry out agent operations.
All changes to the dataset are recorded, to calculate metrics at the end of a simulation run.
\section{Tasks and Fault Definitions}
\label{sec:tasks}

Buildings account for 36\,\% of the global energy use~\cite{IEA} and the top two sinks during operation are HVAC (heating, ventilation, air conditioning) and lighting~\cite{UNEP}.
The tasks currently defined in BOLD focus on lighting. Agents must find and operate the building's lighting system in a fully automated manner in order to minimze electric consumption, thus, in MAS terms, fulfilling simple norms.

We consider different task types with increasing complexity: single-loop tasks and continuous-loop tasks. All tasks are designed for 24-hour runs.
Tasks differ in the number of retrievals and updates they require, as well as in the number of perception-action loops agents must perform.
These dimensions directly echo our discussion on the non-situatedness of Linked Data agents (Sec.~\ref{sec:intro}, items 1 and 3).
Moreover, certain tasks require that agents perform reasoning to make informed decisions (item 2).
Table \ref{tab:task-summary} summarizes the 10 tasks of BOLD.

\begin{table}[bt]
    \caption{BOLD tasks with lower bounds for read/write operations and loops (non-deterministic start/end ($\ast$) or loop count ($\dagger$)) to correctly achieve the task.}
    \label{tab:task-summary}\centering
    \begin{tabular}{l r r r c}
        \toprule
        \textbf{Task} & \textbf{Reads} & \textbf{Writes} & \textbf{Loops} & \textbf{Reasoning} \\
        \midrule
        TS1 & 0 & 146 & 1 & \\
        TS2 & 146 & 146 & 1 & \\
        TS3 & 0 & 6 & 1 & $\checkmark$ \\
        TC1 & 146 & 146 & 2 & \\
        TC2 & 146 & 146 & 2 & $\checkmark$ \\
        \bottomrule
\end{tabular}
    \begin{tabular}{l r r r c}
	\toprule
	\textbf{Task} & \textbf{Reads} & \textbf{Writes} & \textbf{Loops} & \textbf{Reasoning} \\
	\midrule
        TC3 & 147 & 146 & 2$^\ast$ & \\
        TC4 & 128 & 64 & 2$^\ast$ & \\
        TC5 & 128 & 64 & $\sim$4$^{\ast\dagger}$ & \\
        TC6 & 192 & 64 &  $\sim$4$^{\ast\dagger}$ & \\
        TC7 & 256 & 64 &  $\sim$4$^{\ast\dagger}$ & \\
        \bottomrule
    \end{tabular}
\end{table}

\subsection{Single-loop Tasks}

Single-loop tasks require to carry out operations only once in the environment.
Formally, a single-loop task defined by some faults $\Gamma$ on a simulation environment $\langle d, u \rangle$ has the following property: if $\Gamma_{\rho,t} = \emptyset$ for some $t$ and $d_{t+1} \in u(d_t)$ (\ie in the absence of agent operations), then it holds that $\Gamma_{\rho,t+1} = \emptyset$.

As scenario, we consider the simple control of lights, \eg that a janitor would trigger when the whole building closes (TS1) or when they test functionalities of the system (TS2 and TS3).

\begin{description}
\item[TS1]
A light that is on is considered a fault.
This task does not require any data point to be read, the agent must merely find all lights and turn them off.

\item[TS2]
In this task, a light that has not been toggled since the beginning of the run is considered a fault. 
In contrast to TS1, this task involves perception. The agent must first read the state of a luminance command and then toggle.

\item[TS3]
Similar to TS1, but only a subset of the lights has to be switched off, namely those in rooms dedicated to `personal hygiene' (toilet and shower).
Agents must properly classify rooms, which requires that they first read sub-class axioms defined in the environment in a custom RDF vocabulary.
These axioms specify, e.g. that `disabled toilets' are a kind of `toilets', themselves a kind of `personal hygiene' rooms.
As in TS1, a light in a toilet or shower that is on is considered a fault. To ensure agents correctly classify rooms, any light that has been toggled in other rooms is also considered a fault. 

\end{description}

\subsection{Continuous-loop Tasks}

In single-loop tasks, changes in the environment have no effect on success. This is not true of continuous-loop tasks.
In continuous-loop tasks, changes in the environment may cause faults to appear, agents must thus continuously monitor the environment.
The initial value of all lights is randomized so that fault rate always equals 0 in the absence of agent operation, regardless of the task.

Continuous tasks TC1 and TC2 involve time only. TC3 and TC4 involve illuminance sensors. TC5-TC7 introduce occupancy sensors.
\begin{description}
\item[TC1]
In this scenario, a weather report provides sunrise and sunset times.
A light that is off during the day is considered a fault, under the assumption that the building is likely to be occupied. Conversely, lights should be off during the night as no one is expected to be in the building.
An agent needs to retrieve sunrise and sunset timestamps, and turn lights on or off accordingly.
\item[TC2]
In this scenario, in addition to sunrise and sunset times, parts of the building (floors) expose different opening and closing times beyond which no light should be on.
The ground floor is assumed to close later (11pm) than other floors (7pm). All floors open at 8am.
When a floor is open, any light on the floor that is off is considered a fault. 
In this task, the agent must perform automated reasoning to infer what room belongs to each floor in order to decide whether lights in the room should be on or off at a given point in time.
\item[TC3]
In task TC3, a light that is off is a fault if outside illuminance is below a certain threshold (10,000 lux).
In this scenario, we assume the building is equipped with a weather station mounted on its rooftop that includes a light sensor.
By applying a threshold, we want to determine whether the lights in the entire building should be on or off.
In contrast to TC1 and TC2, the times at which light should be on or off is randomly generated to simulate cloud coverage.
Yet, agents could anticipate when to perform an action, as illuminance on the surface of the building varies from sunrise until sunset.
\item[TC4]
In task TC4, a fault is defined as in TC3 but at the level of a single room.
We only consider the rooms in the building that are equipped with luminance sensors so as to decide for each room whether lights should be on or off by applying a global threshold (500 lux).
\item[TC5]
In TC5, a fault is any light that is off while an occupant is detected in the room.
In this scenario, we assume that the rooms in the building are equipped with occupancy sensors.
Using those sensors only, we want to determine whether the lights should be on or off.
The challenge for agents in this task is to continuously monitor occupants coming in and leaving the building in a non-deterministic fashion.
Although the simulated environment displays a clear occupancy pattern throughout the day, an agent cannot know it upfront, and can only build a model of occupancy by repeated observation.
\item[TC6]
This task combines TC5 with the constraints of TC4: a fault is any light that is off while an occupant is detected in the room \emph{and} illuminance is below 500 lux.
In this scenario, we want to raise energy efficiency and only turn on light in rooms with occupants and low illuminance.
An agent faces less potential faults but overcoming them requires a more advanced model of the environment (or faster perception-action loops).
\item[TC7]
TC7, the most challenging task of the benchmark adds one more constraint to TC6: instead of a global illuminance threshold, agents have to make decision based on the preferences of occupants, which they can update at any time during the simulation via \nolinkurl{brick:Setpoint} resources.
In this scenario, we want to raise the individual comfort of occupants. The environment includes a set points for each illuminance sensor in the building. %
Thus, the agent must read the value that was last set by the room's occupant before deciding.
\end{description}

Future variations on TC3-TC7 could be based on temperature instead of illumination, to consider more elaborate control with feedback loops.
Further versions of the benchmark could also consider incomplete environmental descriptions and defective devices.
Agents could e.g. leverage additional topological information to infer probable values for the missing data points.
We believe, however, that the nine BOLD tasks provided here represent a significant challenge to Linked Data agents, which must combine fast retrieval with non-trivial decision making.
\section{Showcase Evaluation}
\label{sec:evaluation}

We now provide an evaluation of the tasks introduced in the previous section, and discuss as the relevance of the metrics we defined for the benchmark.
This showcase evaluation relies on the Linked-Data-Fu engine \cite{DBLP:conf/www/StadtmullerSHS13}, which can perform HTTP operations, do reasoning over Linked Data, and execute rule-based programs~\cite{DBLP:conf/www/KaferH18}. %
Linked-Data-Fu is a multi-threaded engine optimized towards inference and communication.
However, it does not directly provide support for agent-oriented programming features such as planning or memorization.
In the following, we compare the performances of two Linked-Data-Fu agents and show that both features (planning, memorization) are desired for performance improvements.

The two Linked-Data-Fu agents we compare are configured as follows: a first agent (\texttt{ldfu}) is seeded with the building's URI and needs to discover all data points by following links.
The second agent (\texttt{ldfu-pre\-fetch}) behaves as if it had prefetched the model of the building, which allows immediate access to data points.
Both agents execute a program that encodes condition-action rules, where the condition is a fault, as indicated in Table~\ref{tab:task-summary}, and the action is a HTTP \texttt{PUT} request sent to the environment as a fix.
Both agents repeat program execution as fast as possible until a run is over.

Table~\ref{tab:ldfu-eval} summarizes the performance of our two baseline agents against our BOLD simulation server. Experiments were run on a single machine with an Intel Core i5 processor (8GB RAM, 8 cores shared between agent and server processes).
The table features all four metrics presented in Sec~\ref{sec:preliminaries}: fault rate (FR), average fault count (AFC), normalized fault count (NFC) and read/write ratio (RWR).
\begin{table}[t]
\centering
\caption{Performance on single-loop and continuous-loop tasks of a baseline agent.}
\label{tab:ldfu-eval}
\begin{tabular}{lr@{ / }lr@{ / }lr@{ / }lr@{ / }l@{ }c}
\toprule
\multirow{2}{*}{\textbf{Task}} & \multicolumn{8}{c}{\texttt{ldfu} / \texttt{ldfu-prefetch}} \\
& \multicolumn{2}{c}{\textbf{FR}} & \multicolumn{2}{c}{\textbf{AFC}} & \multicolumn{2}{c}{\textbf{NFC}} & \multicolumn{3}{c}{\textbf{RWR}} \\
\midrule
TS1 & 0.04 & 0.03 & 140 & 98 & 0.96 & 0.67 & 19 & 0 & (0) \\
TS2 & 0.09 & 0.05 & 137 & 121 & 0.94 & 0.83 & 19 & 1 & (1) \\
\midrule
TS3 & 0.08 & 0.02 & 6 & 6 & 1 & 1 & 167 & 0 & (0) \\
\midrule
TC1 & 0.12 & 0.03 & 100 & 98 & 0.18 & 0.04 & 458 & 102 & (1) \\
TC2 & 0.13 & 0.05 & 66 & 59 & 0.11 & 0.04 & 540 & 85 & (1) \\
\midrule
TC3 & 0.08 & 0.03 & 61 & 58 & 0.08 & 0.03 & 1831 & 282 & (1) \\
TC4 & 0.23 & 0.09 & 17 & 16 & 0.11 & 0.05 & 983 & 211 & (2) \\
\midrule
TC5 & 0.42 & 0.29 & 16 & 15 & 0.15 & 0.10 & 395 & 68 & (2) \\
TC6 & 0.26 & 0.31 & 11 & 10 & 0.09 & 0.11 & 800 & 105 & (2) \\
TC7 & 0.40 & 0.31 & 16 & 12 & 0.20 & 0.11 & 628 & 127 & (4) \\
\bottomrule
\end{tabular}\\
\scriptsize  FR: fault rate; AFC: average fault count; NFC: normalized fault count;\\RWR: read/write ratio (ideal ratio in parenthesis, calculated from Table~\ref{tab:task-summary}).
\end{table}
As expected, \texttt{ldfu-prefetch}, which is directly provided with information about the building, consistently outperforms \texttt{ldfu}.

Both agents show low fault rates for TS1, TS2 and TS3. As Linked-Data-Fu is optimised for such workloads, we regard those rates as lower bound. Rather, TS1-TS3 can be regarded as test cases to assert the correctness of an implementation.

Results for TC1 and TC2 are also showing satisfactory results regarding FR and NFC. However, the number of reads could be significantly reduced by optimising the agents for the task at hand.
In those two tasks, agents essentially only synchronize with the environment.
Similarly, in TC3, outside illuminance evolves linearly with time. It should thus be straightforward to calculate when a fault will occur and thus reduce the number of reads (see Sec.~\ref{sec:tasks}).

Regarding TC4-TC7, it is worth comparing the two agents.
Although \texttt{ldfu-pre\-fetch} sends  $>4\times$ fewer requests to the server, \texttt{ldfu-pre\-fetch} only gains $<14\%$ performance on fault rate, implying that fetching static data represents a rather low overhead.
In contrast, the overhead related to exchanging dynamic readings is much more significant.
Both agents perform readings on all data points, regardless of past values. For task TC4, for instance, they perform 128 reads which they complete only after the server has executed 10 iterations, on average.
The agent's reaction time is therefore limited because of a high number of reads. Yet, the task only requires a total of $128 \times 2$ reads during the whole run (one for each light after sunrise and before sunset).
To reduce agent-server interactions, an agent can remember past readings for statistical inference (\eg to predict future illuminance levels).
Such optimization is even more crucial for TC5-7, as indicated by significantly higher fault rates for \texttt{ldfu-pre\-fetch} despite low fault counts.

\section{Related Work}
\label{sec:related-work}

Our work is at the intersection of (Read-Write) Linked Data, MAS and building simulation.
While we went to lengths to make realistic assumptions for the simulation, we do not claim to compete with commercial building simulation tools.
We see building automation mostly a vehicle for evaluating efficient MAS architectures and not as an end in itself, similarly as the recent ProcTHOR~\cite{procthor}.%

There is considerable body of work on benchmarking triple stores with a read-only SPARQL interface, e.\,g.\ SP\textsuperscript{2}BE\-NCH~\cite{DBLP:conf/icde/SchmidtHLP09}, BSBM~\cite{DBLP:journals/ijswis/BizerS09}, LUBM~\cite{DBLP:journals/ws/GuoPH05}, and projects such as LDBC~\cite{DBLP:journals/dbsk/BonczFGL013} and Hobbit~\cite{DBLP:journals/ercim/NgomoGF16}.
Yet, BSBM contains a case that writes via SPARQL.
A benchmark that sends read-only SPARQL requests to multiple sources is FedBench~\cite{DBLP:conf/semweb/SchmidtGHLST11}.
Closer to BOLD are approaches that consider the Linked Data interface, \ie dereferencing of URIs in multiple requests:
Hartig et al.\@~\cite{DBLP:conf/semweb/HartigBF09} turned the dateset of BSBM into dereferencable URIs, and DLUBM~\cite{DBLP:conf/otm/KeppmannMH17} distributes the dataset from LUBM.
SolidBench\footnote{\url{https://github.com/SolidBench/SolidBench.js}} serves a static social network via HTTP.
None of these works consider state-changing operations that send RDF via HTTP.
Thus, to our knowledge, BOLD is the first benchmark to provide a dynamic Read-Write Linked Data environment for benchmarking user agents.

The intersection of agent technologies and (Semantic) Web technologies includes:
JASDL~\cite{DBLP:conf/dalt/KlapiscakB08} combines Jason with the facility to process semantic data given in description logics (DL) ontologies.
Using BOLD, we want to address agent technologies that also consider the other aspects of Semantic Web Technologies: interaction in HTTP and hypermedia, and tone down expressive DL reasoning.
REST-A~\cite{DBLP:conf/paams/GouaichB10} is an agent framework providing an abstraction for perception and actions based on REST operations.
Conversely, Ricci et al.\@ built a framework to build environments and agent organizations based on JaCaMo (a mature MAS framework~\cite{boissier_multi-agent_2020}), REST and RDF~\cite{DBLP:conf/atal/RicciCMBBH19}.
Both approaches focus on lowering the engineering effort to design MAS architectures.
MAS engineering tools are complementary to BOLD: REST-A or JaCaMo could serve as the basis for MAS implementations evaluated with BOLD.
BOLD does not evaluate inter-agent communication, which is an orthogonal topic with specialised benchmarks~\cite{DBLP:conf/atal/MuletSA06}.

\section{Conclusion}
\label{sec:conclusion}

We have presented BOLD, a framework next to a benchmark for measuring the performance of Linked Data agents.
BOLD simulates a building, with occupants moving around and changing the lighting systems.
BOLD includes 10 tasks for agents: 3 simple tasks, which require agents to carry out operations only once, and 7 complex tasks, which require the agents to continuously monitor the environment and carry out requests depending on changes in the environment.
We have provided performance measurements of two baseline agents on all tasks and illustrated how more sophisticated agents could show better performances.

We hope that BOLD --and if it's only its execution framework-- fosters research and development in the area of autonomous agents for Read-Write Linked Data.
As such, BOLD has been an environment for ISWC 2021's All-the-Agents Challenge~\cite{DBLP:conf/semweb/2021atac}, sparked discussions at a recent Dagstuhl seminar~\cite{DBLP:journals/dagstuhl-reports/BoissierCHR23}, and served as basis for a transportation showcase~\cite{mosaik-eswc23}.

\smallskip\noindent
\textit{Resource Availability Statement:}
Find the BOLD server source code, the building data (and scripts for its generation), building update and fault queries for the BOLD tasks online\footnote{\url{https://github.com/bold-benchmark/bold-server}}.
Example rules for agents that perform the BOLD tasks can also be found online\footnote{\url{https://github.com/bold-benchmark/bold-agents}}. The runtime for the sample implementation is only available upon request due to its alpha stadium\footnote{\url{http://linked-data-fu.github.io/}}.

\printbibliography

\end{document}